\pdfoutput=1
\documentclass[aps,reprint,twocolumn,10pt,superscriptaddress]{revtex4-2}

\usepackage{amsmath}
\usepackage{amssymb}
\usepackage{bbm}
\usepackage{graphicx}
\usepackage[countmax]{subfloat}
\usepackage{framed}
\usepackage{color}
\usepackage{hyperref}
\usepackage{epstopdf}
\usepackage{dsfont}
\usepackage{amsthm}
\usepackage{wrapfig}
\usepackage{relsize}
\usepackage{bm}
\usepackage[utf8]{inputenc}
\usepackage{enumitem}
\usepackage{soul}

\hypersetup{
	colorlinks=true,       
	linkcolor=blue,        
	citecolor=blue,        
	filecolor=magenta,     
	urlcolor=blue         
}
\usepackage{blindtext}

\begin{document}

\title{Kramers--Kronig relations and precision limits in quantum phase estimation}

\author{Ilaria Gianani}
\affiliation{Dipartimento di Scienze, Universit\'a degli Studi Roma Tre, Via della Vasca Navale, 84, 00146 Rome, Italy}
\author{Marco Barbieri}
\affiliation{Dipartimento di Scienze, Universit\'a degli Studi Roma Tre, Via della Vasca Navale, 84, 00146 Rome, Italy}
\affiliation{Istituto Nazionale di Ottica, CNR, Largo Enrico Fermi 6, 50125 Florence, Italy}
\author{Francesco Albarelli}
\affiliation{Faculty of Physics, University of Warsaw, Pasteura 5, PL-02-093 Warszawa, Poland}
\author{Adriano Verna}
\affiliation{Dipartimento di Scienze, Universit\'a degli Studi Roma Tre, Via della Vasca Navale, 84, 00146 Rome, Italy}
\author{Valeria Cimini}
\affiliation{Dipartimento di Scienze, Universit\'a degli Studi Roma Tre, Via della Vasca Navale, 84, 00146 Rome, Italy}
\affiliation{Dipartimento di Fisica, Sapienza Universit\`{a} di Roma, Piazzale Aldo Moro 5, I-00185 Roma, Italy}
\author{Rafał Demkowicz-Dobrzański}
\affiliation{Faculty of Physics, University of Warsaw, Pasteura 5, PL-02-093 Warszawa, Poland}

\begin{abstract}
The ultimate precision in any measurement is dictated by the physical process implementing the observation.
The methods of quantum metrology have now succeeded in establishing bounds on the achievable precision for phase measurements over noisy channels.
In particular, they demonstrate how the Heisenberg scaling of the precision can not be attained in these conditions.
Here we discuss how the ultimate bound in presence of loss has a physical motivation in the Kramers--Kronig relations and we show how they link the precision on the phase estimation to that on the loss parameter.
\end{abstract}

\maketitle

The purpose of quantum metrology is to establish the best possible strategy for performing the measurement of a set of parameters. This dictates the choice of the state preparation and measurement that, for a given evolution governed by those parameters, is able to deliver the most accurate estimate~\cite{GLMreviewScience,PhysRevLett.96.010401,ParisReview}. Early efforts were prompted by the possibility of improving the scaling of the precision of optical phases thanks to the use of quantum resources. However, this capability is hampered by the presence of loss: more sophisticated tools have been developed in order to investigate non-ideal scenarios in parameter estimation~\cite{davidovich,PhysRevLett.111.120401,Demkowicz-Dobrzanski:2012ly}.
The current goals are set on achieving robust operation of quantum metrological protocols, by producing states that can operate efficiently in the loss channels, as well as under more general noise models, albeit without recovering an improved scaling ~\cite{PhysRevLett.102.040403,PhysRevA.83.063836,PhysRevA.83.021804,konrad,PhysRevLett.106.153603,PhysRevA.97.032305}. As customary in quantum metrology, these studies assume a fixed noisy channel, with a detailed characterisation available before its use for phase estimation. A different approach considers loss as a parameter to be estimated, invoking multiparameter methods~\cite{PhysRevA.89.023845,PhysRevLett.123.200503,PhysRevLett.124.140501,Albarelli2021}, in  either the independent or the correlated case.
While the exact expressions of the achievable precision are often cumbersome, bounds are found which enjoy simpler forms, and are  achievable asymptotically for a large amount of resources~\cite{PhysRevA.83.021804}.  

When investigating material samples, the Kramers--Kronig relations (KKR) impose a more involved inter-dependence between these parameters, as the refractive index at any given frequency can be written by means of an integral of the absorption coefficient over the frequency domain~\cite{alma991015534669703276}. This problem has  a more involved structure, as it correlates a single parameter with a continuum, and its possible implications for metrology have not been explored. In this article we demonstrate how the KKR can be employed to establish connections between the ultimate precision limits which have been found independently for the estimation of phase and absorption. In our investigations, we have made use of concepts taken from quantum function estimation, a novel approach that has been added to the toolbox of quantum metrology in recent years~\cite{PhysRevLett.124.010507}.
Our investigation emphasises how bounds on precision are physically motivated, and descend from fundamental properties. 

Kramers~\cite{Kramers} and Kronig~\cite{deL.Kronig:26} investigated the connection between dispersion and absorption from atomic species due specifically on the dipole emission. This allowed them to establish the relation between real and imaginary part of the complex refractive index $\tilde n(\omega)=n(\omega)+i \kappa(\omega)$ as a function of the optical frequency $\omega$
\begin{equation}
\label{eq:KK}
\begin{aligned}
    n(\omega)=&1+\frac{2}{\pi}\mathcal{P}\int_{0}^{\infty} \frac{\omega'\kappa(\omega')}{\omega'^2-\omega^2}d\omega'\\
    \kappa(\omega)=& -\frac{2\omega}{\pi}\mathcal{P}\int_{0}^{\infty} \frac{n(\omega')}{\omega'^2-\omega^2}d\omega'
    \end{aligned}
\end{equation}
where $\mathcal{P}$ denotes the Cauchy principal value.
At a more fundamental level, similar relations are found between the real and imaginary parts of the linear electric susceptibility~\cite{Landau}, and it is now appreciated how this relation is a manifestation of causality in the linear response of a medium~\cite{alma991015534669703276}. This has made them a precious tool in research in material science. The generality of the KKR has made it possible to find applications well outside their original scope of describing optical response, and these are now employed in acoustics and seismology~\cite{ODonnell,Mikhaltsevitch}, imaging problems~\cite{Horsley:2015yu,Baek:19}, and coherent signal processing~\cite{Mecozzi:16}. 

\begin{figure}[!]
\includegraphics[width=\columnwidth]{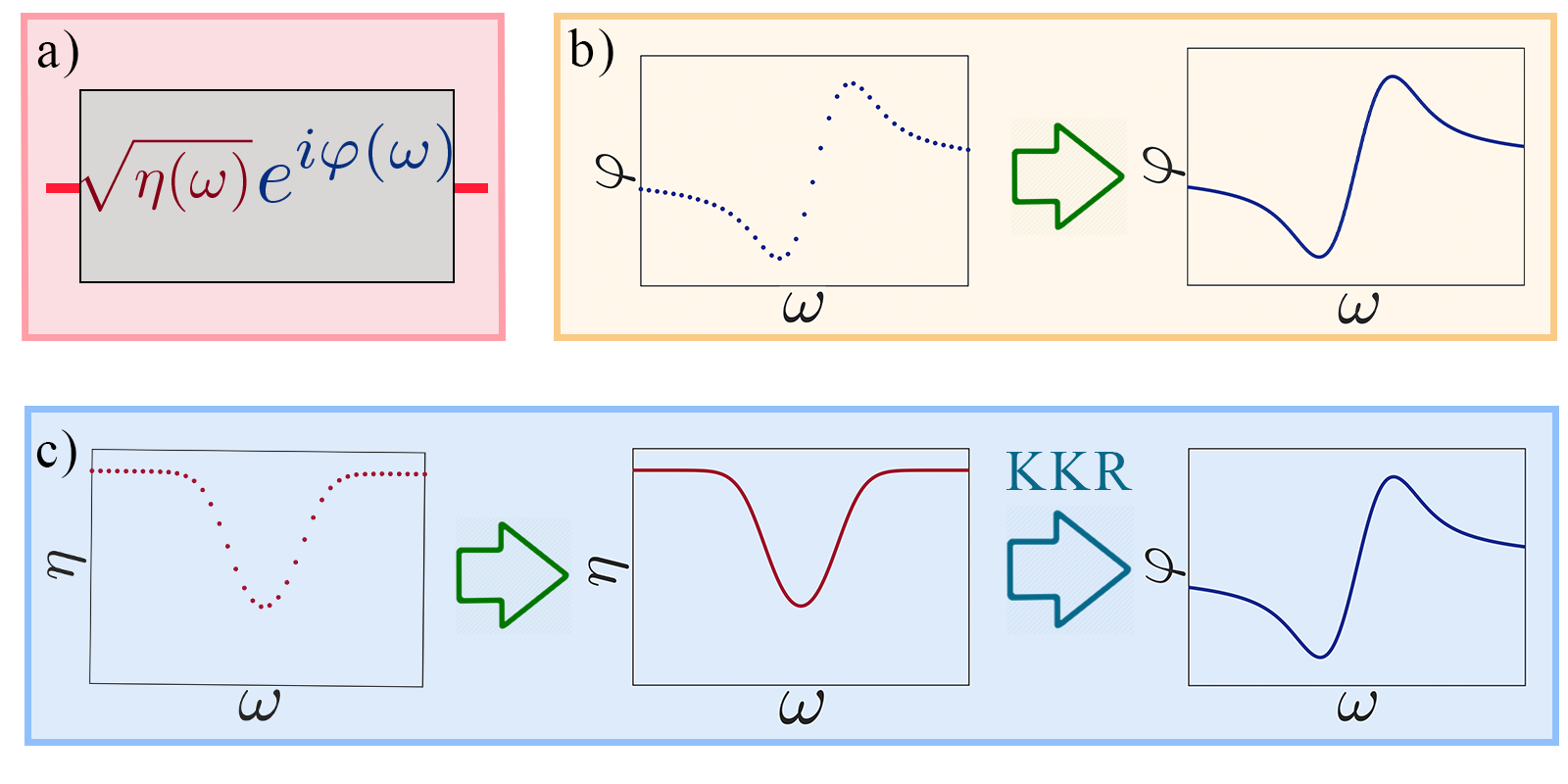}
\caption{Use of the Kramers--Kronig relations for phase reconstruction. {\bf a.} We aim at reconstructing the phase profile of a lossy sample, with frequency-dependent parameters $\varphi(\omega)$ and $\eta(\omega)$. For this, we can either {\bf b.} measure a set of values $\tilde \varphi(\omega_i)$ and interpolate them to obtain the estimated function $\tilde\varphi(\omega)$ or {\bf c.} use the KKR on a reconstructed transmission profile $\tilde \eta(\omega)$. }\label{fig:concept}
\end{figure}

In optics, the relations \eqref{eq:KK} are more conveniently expressed in terms of measurable quantities. When light traverses a length $l$ of this medium, it accumulates in a single pass a phase shift $ \varphi(\omega)=(n(\omega)-1)\omega l/c$, with respect to propagation in the vacuum. The transmission for the intensity follows the Lambert-Beer law:  $\eta(\omega)=e^{-\alpha(\omega)l}=e^{-2\kappa(\omega)\omega l/c}$. Differently from $n(\omega)$ and $\kappa(\omega)$, the quantities $\varphi(\omega)$ and $\eta(\omega)$ can be accessed experimentally. The KKR can be cast in the form   
\begin{equation}
\begin{aligned}
\label{eq:KK1}
    &\varphi(\omega)=-\frac{1}{2\pi}\mathcal{P}\int_{-\infty}^{\infty} \frac{\log \eta(\omega')}{\omega'-\omega}d\omega'= \hat H[\log \sqrt{\eta}](\omega),\\
&\log \sqrt{\eta(\omega)}= -\hat H[\varphi(\omega)],
    \end{aligned}
\end{equation}
where $\hat H$ is the Hilbert transform operator. 
In order to prove these relations, we consider the first of the two equations \eqref{eq:KK}, multiplying both sides by $\omega l/c$:
\begin{equation}
\label{eq:dem}
    \varphi(\omega)=\frac{\omega}{\pi}\mathcal{P}\int_{0}^{\infty}\,d\omega' \frac{\alpha(\omega')l}{\omega'^2-\omega^2}.
\end{equation}
Since the complex refractive index satisfies $n(-\omega)=n^*(\omega)$~\cite{alma991015534669703276}, $\kappa(\omega)$ is an odd function, making $\alpha(\omega)$ an even function. This property can be used to rewrite the integral \eqref{eq:dem} over both negative and positive frequencies; this eventually leads to the first of the equations in \eqref{eq:KK1}. The second is obtained by the properties $\hat H[\hat H[f]]=-f$. Thus, in principle, by measuring the transmission profile $\eta(\omega)$ over the whole frequency spectrum, it is possible to recover the function $\varphi(\omega)$, and viceversa. This form is also the one used in the context of signal processing~\cite{Mecozzi:16}.

In quantum metrology, phase and loss are considered as model examples for the estimation of unitary~\cite{PhysRevLett.96.010401} and dissipative parameters~\cite{PhysRevLett.98.160401,PhysRevA.79.040305}. In ideal conditions, an optical phase can be measured with an uncertainty scaling with the number $p$ of photons in the probe as $1/p^2$, the so called Heisenberg limit~\cite{GLMreviewScience}, as opposed to the optimal scaling with classical light $1/p$. The estimation of phase in a lossy system reveals that, while the Heisenberg limit can not be attained,  a quantum advantage is retained as a constant factor~\cite{Demkowicz-Dobrzanski:2012ly}. Further, studies on the joint estimation of the two parameters have shown that simultaneous optimal estimability can not be implemented~\cite{PhysRevA.89.023845,PhysRevLett.123.200503,Conlon2021}.
While these works have only considered uncorrelated values, in~\cite{PhysRevLett.124.140501}, instead, a form of correlation was introduced, in that the precision limits on a parameter $\chi$, on which both the phase $\varphi(\chi)$ and $\eta(\chi)$ depend, has been investigated. The picture offered by the KKR in the form~\eqref{eq:KK1} is even more involved, as it shows how a single parameter $\varphi_0=\varphi(\omega_0)$ at a given frequency $\omega_0$ is actually related to the whole \emph{function} $\eta(\omega)$. It is then convenient to study the metrological implications of the KKR \eqref{eq:KK1} with the methods the recently introduced quantum function estimation~\cite{PhysRevLett.124.010507,PhysRevA.103.042602}.

We thus define our goal as that of obtaining an estimate $\tilde \varphi(\omega)$ of the actual phase function $\varphi(\omega)$; for quantum metrology, the relevant figure is the error~\cite{PhysRevLett.124.010507}
\begin{equation}
\label{eq:error}
    \delta_\varphi^2=\mathbf{E}\left[\int_{-\infty}^{\infty} |\varphi(\omega)-\tilde\varphi(\omega)|^2 d\omega\right],
\end{equation}
obtained as the expectation value over all the evaluations of $\tilde\varphi(\omega)$. Two routes are followed to obtain such a function, as shown in Fig.\ref{fig:concept}; we first consider the case of a direct measurement of the phase: $\tilde\varphi(\omega)$ is then estimated by assessing the values for some frequencies $\omega_i$, and interpolating in between these points. The error \eqref{eq:error} then originates from two contributions: the statistical uncertainties on the measured phases $\tilde\varphi(\omega_i)$, and the deviations due to the interpolation~\cite{PhysRevLett.124.010507}. The ideal limit has the measured sample dense enough that the second contribution is small with respect to the statistical one, and the total error is thus well approximated by the sum of point-by-point deviations $\Delta\varphi^2(\omega)=\mathbf{E}\left[|\varphi(\omega)-\tilde\varphi(\omega)|^2\right]$, hence
$\delta^2\simeq\int_{-\infty}^{\infty}\Delta\varphi^2(\omega)  d\omega$.

\begin{figure*}[!]
\includegraphics[width=1\textwidth]{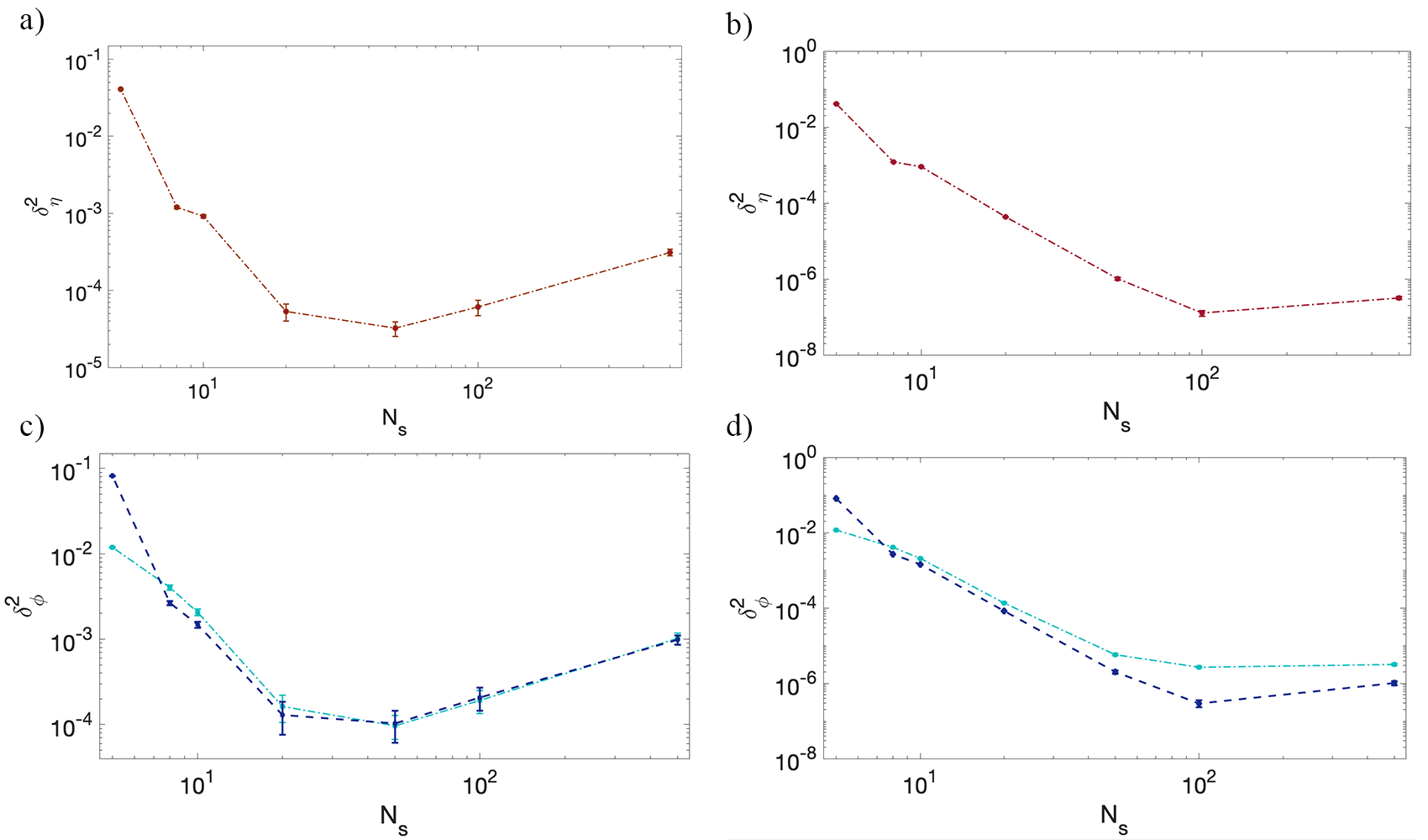}
\caption{Results of the simulated estimation. The error $\delta^2$ is calculated for the transmittivity (red circles) and shown in the top panels a)-b), while different errors for the phase are shown in the bottom panels c)-d): the phase reconstructed through KKR (light blue squares), and  the phase directly estimated at the lossy bound (dark blue diamonds).
The two columns refer to different total resources:  a) and c) $N_{tot}=10^5$, and b) and d) $N_{tot}=10^8$. The error bars are calculated through a Monte Carlo routine based on $N_{mc}=50$ runs, repeating the same numerical experiment.}
\label{fig: results}
\end{figure*}

As an alternative strategy, we can obtain an estimate $\tilde \eta(\omega)$ of the transmission profile, and use this to derive the phase, by means of the KKR \eqref{eq:KK1}. The error on the phase function is then
\begin{equation}
\label{eq:error1}
\begin{aligned}
      \delta_\varphi^2=\frac{1}{4}\mathbf{E}\left[\int_{-\infty}^{\infty} \left\vert\hat H[\log \tilde \eta](\omega)-\hat H[\log \eta](\omega)\right\vert^2 d\omega\right].
\end{aligned}
\end{equation}
We can simplify this expression by invoking the property that, for any square-summable function $f(\omega)$, $\int |f(\omega)|^2d\omega=\int|\hat H[f](\omega)|^2d\omega$: 
\begin{equation}
\label{eq:error1}
        \delta_\varphi^2=\frac{1}{4}\mathbf{E}\left[\int_{-\infty}^{\infty} | \log \tilde \eta(\omega)-\log \eta(\omega)|^2 d\omega.\right]\\
\end{equation}
If the collected sample is sufficiently dense, we can take the same approximation as above, and obtain $\delta_\varphi^2=\int_{-\infty}^{\infty}  \Delta \eta^2(\omega)/(4\eta^2(\omega)) d\omega$, valid when the deviation $\Delta\eta^2 (\omega)=\mathbf{E}\left[|\tilde \eta(\omega)- \eta(\omega)|^2\right]$ is small. The adoption of the KKR thus allows an estimate for the phase function with the the same error we would get from a direct phase measurement with an equivalent uncertainty:
\begin{equation}
\label{eq:dephi}
\Delta\varphi^2_{\rm eq}(\omega)=\frac{\Delta \eta^2 (\omega)}{4\eta^2(\omega)}.
\end{equation}
This quantity, however, should not be interpreted as the error on the individual phase values as obtained from the function $\tilde \varphi(\omega)$.
Optimal estimation of the transmission is obtained by using Fock state probes~\cite{PhysRevA.79.040305} and we allow for the same number of repetitions of the same states for every frequency.
This yields to the uncertainty $\Delta\eta^2(\omega) = (1-\eta(\omega))\eta(\omega)/N$, where, for a $p$-photon state and $M$ repetitions, $N=pM$.
More in general the same form of the optimal precision holds also when substituting $N$ with the average number of photons used to probe the sample at each frequency; in this regard a more practical optimal strategy is to use half of a two-mode squeezed vacuum state~\cite{Nair2018}.
Substituting this expression in \eqref{eq:dephi}, we obtain:
\begin{equation}
\Delta\varphi^2_{\rm eq}(\omega) = \frac{1-\eta(\omega)}{4\eta(\omega)N}.
\label{eq:dephi1}
\end{equation}
This is a known bound on the ultimate quantum limit for lossy phase estimation with $N$ photons~\cite{davidovich,PhysRevA.82.053804} which is asymptotically attainable~\cite{PhysRevA.83.021804}.
Therefore, by using the optimal states for loss estimation, we obtain an estimate $\tilde\varphi(\omega)$ which is, at the leading order, as good as the one we could obtain by employing the optimal states for phase estimation. This discussion, however, comes with the important caveat that the Hilbert transform of a constant function is zero. This implies that the phase profile can be reconstructed from the transmission up to a phase, which needs being fixed upfront. The boundary condition~\cite{alma991015534669703276} that $\phi(\omega)\rightarrow 0$ for $\omega\rightarrow\infty$ may help resolving this ambiguity.

The same reasoning can be applied, in reverse, to the estimation of the transmission $\alpha(\omega)$, obtaining $\delta_{\alpha }^2=4/l^2\int_{-\infty}^{+\infty}\Delta^2\phi(\omega)d\omega$.
If the phase can be estimated optimally, it will asymptotically follow the error~\eqref{eq:dephi1}, thus we obtain $\delta_{\alpha }^2=1/l^2\int_{-\infty}^{+\infty}\frac{1-\eta(\omega)}{ N \eta(\omega)} d\omega$.
As expected, this is equivalent to the error obtained from a direct optimal estimation of $\eta(\omega)$, i.e. $\delta_{\alpha }^2 = \int_{-\infty}^{+\infty} \Delta \alpha^2 (\omega) d\omega$ with $\Delta \alpha^2 (\omega) =  \frac{\Delta^2 \eta (\omega)}{l^2 \eta^2(\omega)} = \frac{1-\eta (\omega)}{l^2 \eta(\omega)} $, since $\alpha = -1/l \log \eta$, by definition.

This reasoning also highlights a different reason why the Heisenberg scaling is lost in lossy phase estimation: were it not so, the equivalent uncertainty $\Delta\alpha^2(\omega)$, and consequently, the  function estimation of $\alpha(\omega)$ would show quantum enhancement in the form of a $1/p^2$ scaling of the uncertainty. However, dissipation is not a coherent process, while, on the other hand, coherent mechanisms are the ones allowing for such an enhancement in the unitary case~\cite{PhysRevLett.96.010401,PhysRevLett.98.160401}. 
If, on the one hand, the KKR are established as a consequence of the physical properties of dispersive and absorptive objects, they can be interpreted as the fact that the complete determination of one function leads to an exhaustive knowledge of the other. Our results interpret this property as optimality in a statistical sense, in that resources are employed for the best use for one parameter as well as the other.

There may occur other mechanisms, including those related to the detectors, which can nevertheless be calibrated in advance. If overall efficiency, without the sample inserted, is $T(\omega)$, the optimal uncertainty on the transmission is $\Delta\eta^2(\omega) = (1-\eta(\omega) T(\omega))T(\omega)/(\eta(\omega) N)$. Inserting this expression in the equivalent phase uncertainty leads to 
$\Delta\varphi^2_{\rm eq}(\omega) = (1-\eta(\omega) T(\omega))/(4\eta(\omega) T(\omega)N)$: the phase profile estimation is influenced by the total transmission $\eta(\omega) T(\omega)$, and the equivalent uncertainty is the optimal permitted by the total loss. The KKR are also useful to link the best precision on phase and transmission allowed by classical light. For this, we use coherent states as benchmarks: loss estimation with the state $\vert \beta \rangle$ leads to the uncertainty $\Delta\eta^2(\omega) = \eta(\omega)/\beta^2$. From this expression, we derive the equivalent uncertainty \eqref{eq:dephi} as $\Delta\varphi^2_{\rm eq}(\omega) = 1/(4\eta(\omega) \beta^2)$, which is the corresponding precision for phase estimation, when loss is considered. In this case too, the KKR relate the classical limits for the individual parameters. Coherent states are indeed optimal among classical states for both phase and loss estimation.

We have explored the implications of the KKR in the practical case, performing numerical calculations. We have considered a Gaussian absorption centered around $\omega_0$ with $\alpha(\omega)=\alpha_0\exp(-(\omega-\omega_0)^2/(2\sigma^2))$. This is a relevant model for those media with an isolated line, or close multiplets, such as quantum memories, whenever the contribution to the phase of other absorption regions can be neglected. This is associated to a phase profile $\varphi(\omega)$ described by a Dawson function~\cite{alma991015534669703276}. The absorption profile has parameters $\alpha_0 l=1$, $\omega_0=0.5$ and $\sigma=0.1$, and the interval $[0,1]$ is considered.

We have simulated the estimation of the functions $\tilde \eta(\omega)$ and $\tilde \phi(\omega)$ from a set of measurements of the transmission using the optimal single-photon states, with a fixed number of total resources $N_{\rm tot}$ divided among $N_s$ sampled points; each of these is then estimated from $N_{\rm ev}=N_{\rm tot}/N_s$ events. For each of the $N_s$ points taken on the transmittivity curve, equally spaced on the $\omega$ axis, the value of $\tilde \eta_i=\tilde \eta(\omega_i)$ is taken from a normal distribution with mean $\eta_i$ and variance $\eta_i(1-\eta_i)/N_{\rm ev}$ - {\it i.e.} we assume an unbiased measurement and a sample sufficiently large to achieve an estimator for $\tilde \eta$ with a Gaussian distribution. A linear interpolation method is employed to connect those points, and obtain a continuous function from these discrete measurements. The error integral on the loss is approximated as a discrete sum  $\delta^2_\eta\simeq \sum_{j=1}^{N_{\rm ref}}|\tilde\eta(\omega_j)-\eta(\omega_j)|^2\delta\omega$. We have defined $N_{\rm ref}=10000$ as the number of interpolated points, $\delta\omega=1/N_{\rm ref}$ as their spacing in frequency. We expect the error to reach a minimum, as a result of the trade off between the resolution and the statistical uncertainty on the measured points~\cite{PhysRevLett.124.010507}. Indeed, for fixed $N_{\rm tot}$, the lower $N_s$, the higher the error from the interpolation, while increasing $N_s$ also increases the statistical contribution to the error~\cite{PhysRevA.103.042602}. Further, we calculate the expected error on $\tilde \varphi(\omega)$ if we were able to estimate phases with a variance given by \eqref{eq:dephi1}.

Starting from $\tilde \eta(\omega)$, the KKR are employed to obtain $\tilde\varphi(\omega)$, which is then compared to the Dawson phase profile. We used the numerical methods introduced in~\cite{ZHOU2009585} for the evaluation of the Hilbert transform, which, for reduced intervals, introduces less numerical errors with respect to the usual approach based on the Fourier Transform~\cite{alma991015534669703276}. Given a function $f(x)$, reconstructed over the interval $[x_{\rm min},
x_{\rm max}]$, its Hilbert transform can be approximated by the sum
\begin{equation}
    H[f](x) \approx \frac{1}{\pi} \sum_{k}f\left(\frac{2k+1}{2^{j+1}}\right)\log\left|\frac{2^jx-k}{2^jx-k-1}\right|, 
\end{equation}
where $j$ is a fixed natural, setting the accuracy of the approximation, and $k$ is an integer satisfying $2^jx_{\rm min}<k<2^jx_{\rm max}$, {\it i.e.} the original interval is sampled in steps of $1/2^j$; for our calculations we used $j=17$, as a compromise between accuracy and computing time.

The results of our simulations are depicted in Figs.~\ref{fig: results}a. and \ref{fig: results}b. Function estimation of $\tilde \eta(\omega)$ and $\tilde \phi(\omega)$ achieve the minimal error for the same number of sampled points $N_s$. For the smaller number of resources, $N_{\rm tot}=1000$, the reconstruction of phase based on the KKR performs close to the limit set by \eqref{eq:dephi1}. When the resources are increased to $N_{\rm tot}=10^8$, a discrepancy appears with respect to the ideal bound. We have verified that this setting reproduced the phase profile reliably when using the transmission sampled with $N_{\rm ref}$ points, with an error of the order $10^{-6}$. Therefore, the saturation of the error observed in Fig.~\ref{fig: results} does originate from the setting of $j$: the statistics of the event is now sufficiently good to reveal systematic effects. We should observe, however, that the uncertainty limit \eqref{eq:dephi1} is optimistic, especially for high values of $\eta(\omega)$; while the bound is asymptotically attainable, convergence to the asymptotic bound is slow~\cite{PhysRevA.83.021804}.

In conclusion, we have shown that by means of the KKR a connection can be established between the ultimate precision limits in loss estimation and lossy phase estimation.
This is remarkable, in that these two limits pertain to rather different experimental arrangements, in terms of probe states and measurements.
Indeed, probe states suited for quantum-enhanced estimation of phase and loss are very different, e.g. Gaussian states squeezed in perpendicular directions.
This means that, for example, the error of a direct estimation of the loss profile using states that are optimal for measuring the phase profile will give a greater error than an indirect estimation via the KKR.
However, coherent states of light are a notable exception: the same precision is obtained with direct or indirect estimation of the two profiles, thanks to the fact that they have equal uncertainty for phase and amplitude quadratures.

The relations \eqref{eq:KK1} demonstrate that the correlation between phase and loss parameters is rather convoluted. Phase actually depends on the loss function on the whole spectrum, thus approaches treating phase and loss values at given frequencies as independent parameters are fully justified~\cite{PhysRevA.83.021804,PhysRevA.89.023845,PhysRevLett.123.200503}. 
Nonetheless, one could perform a joint estimation of both phase and loss; in such a scenario the correct bound on the error on the two profiles should take into account the constraints induced by the KKR. This is a nontrivial problem and an exciting direction for future research.
It should be noticed, however, that, even if one is interested only in phase estimation, loss should be known in order to design the optimal states and for the subsequent data processing.
Under proper conditions, such as for isolated lines, this knowledge can be exploited in order to obtain {\it a priori} probabilities for a Bayesian approach. The presence of nearby spectral features may nevertheless influence the phase profile, inducing systematic effects.

 The possibility of accessing both functions without incurring in a trade-off in the precision suggests that, beyond metrology, quantum resources may have potential  applications in the realisation of Kramers--Kronig coherent receivers~\cite{Mecozzi:16}: this communication protocol reconstructs a complex signal by means of equations similar to \eqref{eq:KK1}---in this instance, the role of the transmittivity is taken by the signal intensity.

The KKR are peculiar to the optical absorption-dispersion mechanism, however, one may attempt generalisations to other contexts, whenever the parameters to be estimated can be traced back to the response function to an applied field. These could then provide insight on the physical origin of the bounds on precision.

\emph{Acknowledgements.}
This work was supported by the European Commission through the FET-OPEN-RIA project STORMYTUNE (G.A. number 899587).
FA and RDD are supported by the National Science Center (Poland) grant No.\ 2016/22/E/ST2/00559.

\bibliography{biblio.bib}

\end{document}